\documentclass[10pt]{article}

\usepackage[a4paper,top=2.5cm,bottom=2.5cm,left=2.5cm,right=2.5cm]{geometry}
\usepackage{times}
\usepackage{natbib}
\usepackage{amssymb}
\usepackage{amsmath}
\usepackage{amsfonts}
\usepackage{epsfig}
\usepackage{graphicx}
\usepackage[font=small]{caption}

\DeclareMathOperator{\Ima}{Im}
\DeclareMathOperator{\Int}{int}

\begin{document}

\title{Stopping of ions in a plasma irradiated by an intense laser field}

\author{H.B. NERSISYAN$^{1,2}$\thanks{E-mail: hrachya@irphe.am}
\textsc{and} C. DEUTSCH$^{3}$ \\ \small{$^{1}$Institute of Radiophysics
and Electronics, 0203 Ashtarak, Armenia} \\
\small{$^{2}$Centre of Strong Fields Physics, Yerevan State University, Alex Manoogian
str. 1, 0025 Yerevan, Armenia} \\
\small{$^{3}$LPGP (UMR-CNRS 8578), Universit\'{e} Paris XI, 91405 Orsay, France}}

\date{\today}

\maketitle

\begin{abstract}
The inelastic interaction between heavy ions and an electron plasma in the presence of an intense
radiation field (RF) is investigated. The stopping power of the test ion averaged with a period of
the RF has been calculated assuming that $\omega _{0} >\omega_{p}$, where $\omega_{0}$ is the
frequency of the RF and $\omega_{p}$ is the plasma frequency. In order to highlight the effect of
the radiation field we present a comparison of our analytical and numerical results obtained for
nonzero RF with those for vanishing RF. It has been shown that the RF may strongly reduce the mean
energy loss for slow ions while increasing it at high--velocities. Moreover, it has been shown, that
acceleration of the projectile ion due to the RF is expected at high--velocities and in the high--intensity
limit of the RF, when the quiver velocity of the plasma electrons exceeds the ion velocity.
\end{abstract}


\section{Introduction}
\label{sec:Intr}

The interaction of charged particles with a plasma in the presence of radiation field (RF) has been a subject
of great activity, starting with the work of Tavdgiridze, Aliev, Gorbunov and other authors \citep{tav70,ali71,ari89,ako97,ner99}.
A comprehensive treatment of the quantities related to inelastic particle--solid and particle--plasma interactions,
like scattering rates and differential and total mean free paths and energy losses, can be formulated in terms
of the dielectric response function obtained from the electron gas model. The results have important applications
in radiation and solid--state physics \citep{rit75,tun77,ech87}, and more recently, in studies of energy deposition
by ion beams in inertial confinement fusion (ICF) targets \citep{ari81,meh81,may82,ari87,ava93,cou94}. On the
other hand, the achievement of high--intensity laser beams with frequencies ranging between the infrared and
vacuum--ultraviolet region has given rise to the possibility of new studies of interaction processes, such as
electron--atom scattering in laser fields \citep{kro73,wei77,wei83}, multiphoton ionization \citep{lom76,bal81},
inverse bremsstrahlung and plasma heating \citep{see73,kim79,lim79}, screening breakdown \citep{mir05}, and other
processes of interest for applications in optics, solid--state, and fusion research. In addition, a promising ICF
scheme has been recently proposed \citep{sto96,rot01}, in which the plasma target is irradiated
simultaneously by intense laser and ion beams. Within this scheme several experiments \citep{fra10,hof10} have been
performed to investigate the interactions of heavy ion and laser beams with plasma targets. An important
aspect of these experiments is the energy loss measurements for the ions in a wide-range of plasma parameters.
It is expected in such experiments that the ion propagation would be essentially affected by the parametric
excitation of the plasma target by means of laser irradiation. This effect has been supported recently by
particle-in-cell (PIC) numerical simulations \citep{hu11}.

In this paper we present a study of the effects of intense RF on the interaction of nonrelativistic projectile ions
with an electron plasma. Our objective is to study two regimes of the ion energy loss which have not been considered
in detail. For the first part of our study, we consider energy loss of a slow ion. In particular, this is
motivated by the fact that the alpha-particles resulting from the nuclear fusion in a very dense plasma with temperature
in the keV range, display a velocity mostly below electron thermal velocity. The second objective of our study is to
investigate the energy loss in high--velocity regime. Previously this has been done for a classical plasma
\citep{tav70,ali71,ner99} treating only the collective excitations as well as in the range of solid--state densities
(fully degenerate plasma) and at the intermediate intensities of the RF \citep{ari89} when the electron quiver amplitude
is comparable to the screening length of the target. To gain more insight into the RF effect on the energy loss process
we consider here the regime of intense RF when the quiver amplitude largely exceeding the typical screening length of the
fully degenerate electron plasma.

The plan of the paper is as follows. In Sec.~\ref{sec:1} we briefly outline the RPA formulation for the energy loss of a
heavy ion uniformly moving in a plasma in the presence of an intense RF. The limiting case of a weak RF is also considered.
In Secs.~\ref{sec:2} and \ref{sec:3} we have calculated the effects of the RF on the mean energy loss (stopping power) of
the test ion considering two somewhat distinct cases with slow (Sec.~\ref{sec:2}) and fast (Sec.~\ref{sec:3}) projectiles
moving in a classical and fully degenerated electron gas, respectively. In the latter case the degenerated electron gas is
treated within a simple plasmon--pole approximation proposed by Basbas and Ritchie \citep{bas82}. It has been shown, that
besides usual stopping in a plasma it is possible to accelerate the charged particles beam through RF. This effect is
expected for fast projectiles and in the high--intensity limit of the RF, when the ''quiver velocity'' of the plasma
electrons exceeds the projectile ion velocity. The results are summarized in Sec.~\ref{sec:sum} which also includes discussion
and outlook.

\section{RPA formulation}
\label{sec:1}

The whole interaction process of the projectile ion with a plasma involves the energy loss and the charge states of the
ion and -- as an additional aspect -- the ionization and recombination of the ion driven by the RF and the collisions
with the plasma particles. A complete description of the interaction of the ion requires a simultaneous treatment of all
these effects including, in particular, the effect of the ion charge equilibration on the energy loss process. In this
paper we do not discuss the charge state evolution of the projectiles under study, but concentrate on the RF effects on
the energy loss process assuming an equilibrium charge state of the ion with an effective charge $Ze$. This is motivated
by the fact that the charge equilibration occurs in time scales which are usually much smaller than the time of passage
of the ion through target.

The problem is formulated using the random--phase approximation (RPA), and includes the effects
of the RF in a self--consistent way. The electromagnetic field is treated in the long--wavelength limit, and the
electrons are considered nonrelativistic. These are good approximations provided that (1) the wavelength of the RF
($\lambda_{0}=2\pi c/\omega_{0}$) is much larger than the typical screening length ($\lambda_{s}=v_{s}/\omega_{p}$
with $v_{s}$ the mean velocity of the electrons and $\omega_{p}$ the plasma frequency), and (2) the ''quiver velocity''
of the electrons in the RF ($v_{E}=eE_{0}/m\omega_{0} $) is much smaller than the speed of light $c$. These conditions
can be alternatively written as (1) $\omega_{0}/\omega_{p} \ll 2\pi c/v_{s}$, (2) $W_{L} \ll\frac{1}{2} n_{0}c (mc^{2})%
(\omega_{0}/\omega_{p})^{2}$, where $W_{L}=cE_{0}^{2}/8\pi$ is the RF intensity. As an estimate in the case of dense
gaseous plasma, with electron density $n_{0} =10^{18}$ cm$^{-3}$, we get $\frac{1}{2} n_{0} mc^{3} \simeq 1.2\times%
10^{15}$ W/cm$^{2}$. Thus the limits (1) and (2) are well above the values obtained with currently available high--power
RF sources, and so the approximations are well justified.

We consider the time--dependent Hamiltonian for the plasma electrons in the presence of both a radiation field (RF)
with vector potential $\mathbf{A}(t)=(c/\omega _{0})\mathbf{E}_{0}\cos (\omega _{0}t)$, and a self--consistent
scalar potential $\varphi (\mathbf{r},t)$ \citep{ari89,ner99}, i.e.,
\begin{equation}
H(t)=\sum_{\mathbf{p}}\frac{1}{2m}\left( \mathbf{p}-\frac{e}{c}\mathbf{A}%
(t)\right) ^{2}c_{\mathbf{p}}^{+}c_{\mathbf{p}}-e\sum_{\mathbf{p,k}}\varphi (%
\mathbf{k},t)c_{\mathbf{p+k}}^{+}c_{\mathbf{p}} ,
\label{eq:1}
\end{equation}%
where $c_{\mathbf{p}}$, $c_{\mathbf{p}}^{+}$ are annihilation and creation operators for electrons with momentum
$\mathbf{p}$, respectively, and $\varphi (\mathbf{k},t)$ is the Fourier transform of $\varphi (\mathbf{r},t)$.

The potential $\varphi (\mathbf{k},t)$ is produced by the external charge and by the induced electronic density, \emph{viz}.,
\begin{equation}
k^{2}\varphi (\mathbf{k},t)=4\pi \rho _{0}(\mathbf{k},t)-4\pi e\sum_{\mathbf{%
p}}N_{\mathbf{p}}(\mathbf{k},t)
\label{eq:2}
\end{equation}%
being $\rho _{0}(\mathbf{k},t)$ the Fourier transform of the external charge density $\rho _{0}(\mathbf{r},t)$, and
$N_{\mathbf{p}}(\mathbf{k},t)= (c_{\mathbf{p-k}}^{+}c_{\mathbf{p}})_{t}$ is the electrons number operator.

The time evolution of the operator $N_{\mathbf{p}}(\mathbf{k},t)$ is determined by the equation
\begin{equation}
i\hbar \frac{\partial N_{\mathbf{p}}(\mathbf{k},t)}{\partial t}=\left[ N_{%
\mathbf{p}}(\mathbf{k},t),H(t)\right] .
\label{eq:3}
\end{equation}%
In particular, for an oscillatory field $\mathbf{A}(t)$ and within random--phase approximation Eq.~\eqref{eq:3} has
the solution \citep{ari89,ner99}
\begin{eqnarray}
&&N_{\mathbf{p}}(\mathbf{k},t)=\frac{ie}{\hbar }\left( f_{\mathbf{p-k}}-f_{%
\mathbf{p}}\right) \int_{-\infty }^{t}dt^{\prime }\varphi (\mathbf{k}%
,t^{\prime })\exp \left[ \frac{i}{\hbar }\left( \varepsilon _{\mathbf{p-k}%
}-\varepsilon _{\mathbf{p}}\right) (t-t^{\prime })\right]  \label{eq:4} \\
&&\times \exp \left[ -i\zeta \left( \sin (\omega _{0}t)-\sin (\omega
_{0}t^{\prime })\right) \right] ,  \nonumber
\end{eqnarray}%
where $\zeta =\mathbf{k\cdot a}$, $\mathbf{a}=e\mathbf{E}_{0}/m\omega _{0}^{2}$ is the oscillation amplitude
of the electrons driven by the RF (quiver amplitude), $\varepsilon _{\mathbf{p}}=p^{2}/2m$ is the electron
energy with momentum $\mathbf{p}$. Here $f_{\mathbf{p}}$ is the equilibrium distribution function for the
electron plasma.

Finally, using Eq.~\eqref{eq:2} and making a further Fourier transformation we obtain a solution for the potential
$\varphi $ in the form
\begin{equation}
\widetilde{\varphi }(\mathbf{k},\omega )=\frac{4\pi \widetilde{\rho }_{0}(%
\mathbf{k},\omega )}{k^{2}\varepsilon (k,\omega )} ,
\label{eq:5}
\end{equation}%
where we have introduced the frequency transforms $\widetilde{\varphi }(\mathbf{k},\omega )$, $\widetilde{\rho %
}_{0}(\mathbf{k},\omega )$ of the quantities
\begin{equation}
\left(
\begin{array}{c}
\widetilde{\rho }_{0}(\mathbf{k},t) \\
\widetilde{\varphi }(\mathbf{k},t)%
\end{array}%
\right) =\left(
\begin{array}{c}
\rho _{0}(\mathbf{k},t) \\
\varphi (\mathbf{k},t)%
\end{array}%
\right) e^{i\zeta \sin (\omega _{0}t)} ,
\label{eq:6}
\end{equation}%
and $\varepsilon (k,\omega )$ is the RPA dielectric function \citep{lin54,lin64}.

We consider a heavy point--like particle with mass $M$ and effective charge $Ze$ which moves with rectilinear trajectory
with constant velocity $\mathbf{v}$. We thus neglect the effect of the RF on the particle assuming that the quiver velocity
of the ion in the laser field $v_{q} =ZeE_{0}/M\omega_{0}\ll v_{s},v$. Here $v_{s}$ is the mean velocity of the target
electrons. The charge density of the point--like ion is then given by $\rho _{0}(\mathbf{r},t)=Ze\delta (\mathbf{r}-\mathbf{v}t)$.
Inserting the Fourier transformation of this formula with respect to $\mathbf{r}$ into Eq.~\eqref{eq:6} and making a further
Fourier transformation we obtain
\begin{equation}
\widetilde{\rho }_{0}(\mathbf{k},\omega )=2\pi Ze\sum_{n=-\infty }^{\infty
}J_{n}(\zeta )\delta \left( \omega -\mathbf{k}\cdot \mathbf{v}+n\omega
_{0}\right) ,
\label{eq:7}
\end{equation}%
where $J_{n}$ is the Bessel function of $n$th order. Using Eqs.~\eqref{eq:5}--\eqref{eq:7} for the self--consistent
potential $\varphi (\mathbf{r},t)$ we finally arrive at
\begin{equation}
\varphi (\mathbf{r},t)=\frac{Ze}{2\pi ^{2}}\sum_{m,n=-\infty }^{\infty
}e^{i(n-m)\omega _{0}t}\int d\mathbf{k}\frac{e^{i\mathbf{k}\cdot (\mathbf{r}-%
\mathbf{v}t)}J_{m}\left( \zeta \right) J_{n}\left( \zeta \right) }{%
k^{2}\varepsilon (k,\mathbf{k}\cdot \mathbf{v}-n\omega _{0})}.
\label{eq:8}
\end{equation}%
This result represents the dynamical response of the medium to the motion of
the test particle in the presence of the RF; it takes the form of an
expansion over all the harmonics of the field frequency, with coefficients $%
J_{n}(\zeta )$ that depend on the intensity $W_{L}\propto a^{2}$.

From Eq.~\eqref{eq:8} it is straightforward to calculate the electric field $%
\mathbf{E}(\mathbf{r},t)=-\nabla \varphi (\mathbf{r},t)$, and the time
average (with respect to the period $2\pi /\omega _{0}$ of the laser field)
of the stopping field $\mathbf{E}_{\mathrm{stop}}=\langle \mathbf{E}(\mathbf{%
v}t,t)\rangle $ acting on the particle. Then, the averaged stopping power (SP) of the test
particle becomes
\begin{equation}
S\equiv -Ze\frac{\mathbf{v}}{v}\cdot \mathbf{E}_{\mathrm{stop}}=\frac{%
2Z^{2}e^{2}}{(2\pi )^{2}v}\sum_{n=-\infty }^{\infty }\int d\mathbf{k}\frac{%
\mathbf{k}\cdot \mathbf{v}}{k^{2}}J_{n}^{2}(\zeta )\Ima\frac{-1}{\varepsilon
(k,\Omega _{n}(\mathbf{k}))}
\label{eq:9}
\end{equation}%
with $\Omega _{n}(\mathbf{k})=n\omega _{0} +\mathbf{k}\cdot \mathbf{v}$.

To illustrate the effects of the RF it is convenient to take into account the symmetry of the integrand
in Eq.~\eqref{eq:9}, with respect to the change $\mathbf{k},n\to -\mathbf{k},-n$. Using also the property
of Bessel functions, $J_{-n}^{2}(\zeta )=J_{n}^{2}(\zeta )$, we obtain
\begin{equation}
S=\frac{Z^{2}e^{2}}{2\pi ^{2}v}\int d\mathbf{k}\frac{\mathbf{k}\cdot
\mathbf{v}}{k^{2}}\left[ J_{0}^{2}(\zeta )\Ima\frac{-1}{\varepsilon (k,\mathbf{k}\cdot \mathbf{v})}%
+2\sum_{n=1}^{\infty }J_{n}^{2}(\zeta )\Ima\frac{-1}{\varepsilon (k,\Omega _{n}(\mathbf{k}))}\right] .
\label{eq:9a}
\end{equation}
Hence, the SP depends on the particle velocity $\mathbf{v}$, the frequency $\omega _{0}$ and the intensity
$W_{L}=cE_{0}^{2}/8\pi $ of the RF (the intensity dependence is given through the quiver amplitude $\mathbf{a}$).
Moreover, since the vector $\mathbf{k}$ in Eq.~\eqref{eq:9a} is spherically integrated, $S$ becomes also a
function of the angle $\vartheta $ between the velocity $\mathbf{v}$, and the direction of polarization of RF,
represented by $\mathbf{a}$.

By comparison, the SP in the absence of the RF is given by \citep{deu86,pet91}
\begin{equation}
S_{B}=\frac{Z^{2}e^{2}}{2\pi ^{2}v}\int d\mathbf{k}\frac{\mathbf{k}\cdot
\mathbf{v}}{k^{2}}\Ima\frac{-1}{\varepsilon \left(k,\mathbf{k}\cdot \mathbf{v}\right) } .
\label{eq:9b}
\end{equation}%

In the presence of the RF the SP $S_{B}$ is modified and is given by the first term in Eq.~\eqref{eq:9a}
(\textquotedblright no photon\textquotedblright\ SP)
\begin{equation}
S_{0}=\frac{Z^{2}e^{2}}{2\pi ^{2}v}\int d\mathbf{k}\frac{\mathbf{k}\cdot
\mathbf{v}}{k^{2}}J_{0}^{2}(\zeta )\Ima\frac{-1}{\varepsilon \left(k,\mathbf{k}\cdot \mathbf{v}\right) } .
\label{eq:9c}
\end{equation}

Next we consider the case of a weak radiation field ($a<\lambda_{s}$, where $\lambda_{s}$ is the characteristic
screening length) at arbitrary angle $\vartheta $ between $\mathbf{v}$ and $\mathbf{E}_{0}$. In Eq.~\eqref{eq:9a}
we keep only the quadratic terms with respect to the quantity $\mathbf{a}$ and for the stopping power $S$ we obtain
\begin{equation}
S =S_{B}+\frac{Z^{2}e^{2}}{4\pi ^{2}v}\int \frac{d\mathbf{k}}{k^{2}} (\mathbf{k}\cdot
\mathbf{v})(\mathbf{k}\cdot \mathbf{a})^{2} \Ima\left[ \frac{1}{%
\varepsilon (k,\omega _{0}+\mathbf{k}\cdot \mathbf{v})}-\frac{1}{\varepsilon
(k,\mathbf{k}\cdot \mathbf{v})}\right] ,
\label{eq:25}
\end{equation}%
where $S_{B}$ is the field-free SP given by Eq.~\eqref{eq:9b}. Note that due to the isotropy of the
dielectric function $\varepsilon (k,\omega)$ the angular integrations in Eqs.~\eqref{eq:9a}--\eqref{eq:25}
can be easily done.

It is well known that within classical description an upper cutoff parameter $k_{\max }=1/r_{\min }$ (where $r_{\min }$
is the effective minimum impact parameter) must be introduced in Eqs.~\eqref{eq:9b} and \eqref{eq:25} to avoid the
logarithmic divergence at large $k$. This divergence corresponds to the incapability of the classical perturbation
theory to treat close encounters between the projectile particle and the plasma electrons properly. For $r_{\min }$
we use the effective minimum impact parameter excluding hard Coulomb collisions with a scattering angle larger than
$\pi /2$. The resulting cutoff parameter $k_{\max } \simeq m(v^{2} +v_{\mathrm{th}}^{2})/|Z|e^{2}$ is well known for
energy loss calculations (see, e.g., \citet{zwi99,ner07} and references therein). Here $v_{\mathrm{th}}$ is the thermal
velocity of the electrons. In particular, at low projectile velocities this cutoff parameter reads $k_{\max }=T/|Z|e^{2}$,
where $T$ is the plasma temperature given in energy units.

\section{Energy loss of slow ions}
\label{sec:2}

In this section subsequent derivations are performed for the classical plasma and in the low--velocity limit
of the ion. In this case the RPA dielectric function is given by \citep{fri61}
\begin{equation}
\varepsilon (k,\omega )=1+\frac{1}{k^{2}\lambda _{\mathrm{D}}^{2}}W\left(
\frac{\omega }{kv_{\mathrm{th}}}\right) ,
\label{eq:10}
\end{equation}%
where $\lambda _{\mathrm{D}}$ is the Debye screening length, and $W(z)=g(z)+if(z)$ is the plasma dispersion function
\citep{fri61} with
\begin{equation}
g(z)=1- z e^{-z^{2}/2}\int_{0}^{z}e^{t^{2}/2}dt , \quad
f(z)=\sqrt{\frac{\pi }{2}}ze^{-z^{2}/2} .
\label{eq:12}
\end{equation}%

Consider now the SP determined by Eq.~\eqref{eq:9a} in the limit of low--velocities, when $v\ll v_{\mathrm{th}}$.
As discussed above we also assume that $v\gg v_{q}$ and neglect the effect of the RF on the ion. In the limit of
the low--velocities from Eqs. \eqref{eq:9a}--\eqref{eq:12} we obtain
\begin{equation}
S(\gamma ,a,\vartheta )=S_{B}\Xi (\gamma ,a,\vartheta ) ,
\label{eq:14}
\end{equation}%
where
\begin{equation}
\Xi (\gamma ,a,\vartheta )=\Xi _{1}(\gamma ,a)+\Xi _{2}(\gamma ,a)\sin ^{2}\vartheta ,
\label{eq:15}
\end{equation}%
\begin{eqnarray}
&&\Xi _{s}(\gamma ,a) =\frac{6}{\psi (\xi )}\left\{ \int_{0}^{\xi }\frac{k^{3}dk}{%
(k^{2}+1)^{2}}\int_{0}^{1} J_{0}^{2}(Ak\mu ) f_{s}(\mu ) d\mu \right.  \label{eq:16} \\
&&\left. +2\sqrt{\frac{2}{\pi }}\sum_{n=1}^{\infty }\int_{0}^{\xi }\Ima\left[
\frac{W_{1}(n/k\gamma )k^{3}dk}{(k^{2}+W(n/k\gamma ))^{2}}\right]
\int_{0}^{1} J_{n}^{2}(Ak\mu ) f_{s}(\mu ) d\mu \right\} \;.  \nonumber
\end{eqnarray}%
Here $s=1,2$, and $f_{1}(\mu )=\mu^{2}$, $f_{2}(\mu )=\frac{1}{2}(1-3\mu^{2})$. Note that at the absence of
the laser field (i.e., at $a\to 0$) $\Xi_{1}(\gamma ,a)\to 1$, $\Xi _{2}(\gamma ,a)\to 0$. In this case the
SP is determined by the quantity $S_{B}$ in Eq.~\eqref{eq:9b} \citep{deu86,pet91}
\begin{equation}
S_{B}=\sqrt{\frac{2}{\pi }}\frac{Z^{2}e^{2}}{6\lambda _{\mathrm{D}}^{2}}%
\frac{v}{v_{\mathrm{th}}}\psi (\xi ) ,
\label{eq:17}
\end{equation}%
where
\begin{equation}
\psi (\xi )=\ln (1+\xi ^{2})-\frac{\xi ^{2}}{1+\xi ^{2}}
\label{eq:18}
\end{equation}%
is the Coulomb logarithm with $\xi =k_{\max }\lambda _{\mathrm{D}}$. Also in
Eqs.~\eqref{eq:14}--\eqref{eq:16} we have introduced the angle $\vartheta $ between the velocity $\mathbf{v}$
and the polarization $\mathbf{a}$ vectors, $W_{1}(z)=dW(z)/dz$, $A=a/\lambda _{D}$, $\gamma =\omega _{p}/%
\omega _{0}<1$. Note that while the $k$ integral in Eq.~\eqref{eq:9b} diverges logarithmically in a field--free
case, Eqs.~\eqref{eq:9c} and \eqref{eq:16} are finite and do not require any cutoff. The Bessel functions
involved in these expressions due to the radiation field guarantee the convergence of the $k$--integrations.
However, since in the sequel we shall compare Eqs.~\eqref{eq:14}--\eqref{eq:16} with field--free SP $S_{B}$, for
consistency the upper limits of the $k$--integrals in Eq.~\eqref{eq:16} are kept finite with the same upper
cutoff parameter as in Eqs.~\eqref{eq:9b} and \eqref{eq:17}.

\begin{figure}[tbp]
\centering{
\includegraphics[width=80mm]{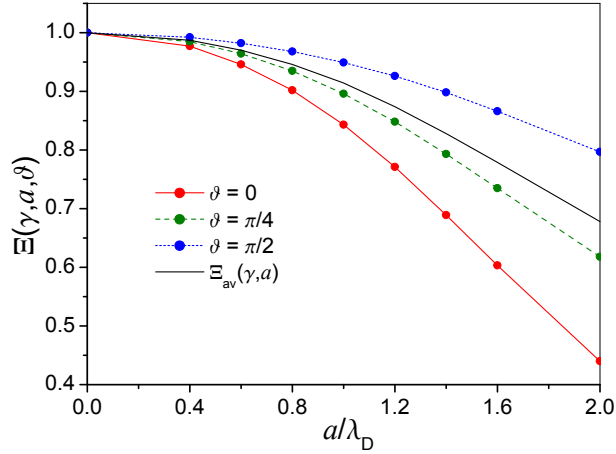}}
\caption{The dimensionless quantities $\Xi (\gamma ,a,\vartheta )$ (the lines with symbols) and
$\Xi_{\mathrm{av}} (\gamma ,a)$ (the solid line without symbols) vs the intensity parameter
of the laser field $a/\lambda _{\mathrm{D}}$ for $\vartheta =0$ (solid line), $\vartheta =\pi /4$
(dashed line), $\vartheta =\pi /2$ (dotted line) and for $\omega _{0}=1.2\omega _{p}$.}
\label{fig:1}
\end{figure}

In many experimental situations, the ions move in a plasma with random orientations of $\vartheta $
with respect to the direction of the polarization of laser field $\mathbf{a}$. The stopping power
appropriate to this situation may be obtained by carrying out a spherical average over $\vartheta$
of $S(\gamma ,a,\vartheta )$ in Eqs.~\eqref{eq:14} and \eqref{eq:15}. We find
\begin{equation}
S_{\mathrm{av}}(\gamma ,a ) =S_{B}\left[ \Xi _{1}(\gamma ,a)+\frac{2}{3}\Xi
_{2}(\gamma ,a)\right] \equiv S_{B} \Xi_{\mathrm{av}} (\gamma ,a) .
\label{eq:19}
\end{equation}

The study of the effect of a radiation field on the SP is easier in the case
of low-intensities $W_{L}$ when $a<\lambda _{\mathrm{D}}$. Then considering
in Eqs.~\eqref{eq:14}--\eqref{eq:16} only the quadratic terms with respect
to $a$ for the SP $S(\gamma ,a,\vartheta )$ we obtain
\begin{equation}
S(\gamma ,a,\vartheta )=S_{B}\left[ 1-\frac{a^{2}}{5\lambda _{\mathrm{D}}^{2}%
}(2\cos ^{2}\vartheta +1)D(\gamma ,\xi )\right] ,
\label{eq:20}
\end{equation}%
where
\begin{equation}
D(\gamma ,\xi )=\frac{1}{\psi (\xi )}\int_{1/\xi }^{\infty }\frac{dx}{x^{3}}%
\left\{ \frac{1}{(x^{2}+1)^{2}}-\sqrt{\frac{2}{\pi }}\Ima\left[ \frac{%
W_{1}(x/\gamma )}{\left( 1+x^{2}W(x/\gamma )\right) ^{2}}\right] \right\} .
\label{eq:21}
\end{equation}%
Taking into account that $\gamma <1$ and $\xi \gg 1$ from Eqs.~\eqref{eq:20} and \eqref{eq:21} we finally
obtain $D(\gamma ,\xi )\simeq 3/4\gamma ^{2}$. It is seen that at low--velocities the SP $S(\gamma ,a,\vartheta )$
decreases with the intensity of radiation field.

In Figure~\ref{fig:1} the quantities $\Xi (\gamma ,a,\vartheta )$ and $\Xi_{\mathrm{av}} (\gamma ,a)$ are shown vs the
intensity parameter $a/\lambda _{\mathrm{D}}$ of the laser field for three values of angles $\vartheta =0$,
$\vartheta =\pi /4$, $\vartheta =\pi /2$ and for $\omega _{0}=1.2\omega _{p}$. It is convenient to represent
the intensity parameter $a/\lambda _{\mathrm{D}}$ in the form $a/\lambda _{\mathrm{D}}=0.18\lambda_{0}^{2}%
\sqrt{n_{0}W_{L}/T}$, where the wavelength ($\lambda _{0}$) and the intensity ($W_{L}$) of the laser field
and the density ($n_{0}$) and the temperature ($T$) of plasma are measured in units $\mu $m, $10^{15}$ W/cm$^{2}$,
$10^{20}$ cm$^{-3}$ and keV, respectively. As an example consider the case when the electron quiver amplitude
reaches the Debye screening length, $a=\lambda_{\mathrm{D}}$. For the values of the RF and plasma parameters
with $\lambda_{0}=0.5$~$\mu$m, $n_{0}=10^{18}$~cm$^{-3}$, $T=0.1$~keV, the above condition is fulfilled at
the radiation field intensity $W_{L}=4.94\times 10^{18}$~W/cm$^{2}$.

From Figure~\ref{fig:1} it is seen that the intense laser field may strongly reduce the SP of the low--velocity
ion. And as expected the effect of the radiation field is maximal for $\vartheta =0$. Note that in this case
and at $a=\lambda_{\mathrm{D}}$ the radiation field reduces the energy loss $S_{B}$ approximately by 15 \%.
For explanation of the obtained result let us consider a simple physical model. The stopping power of the ion
is defined as $S=-(1/v)\langle dW/dt\rangle $, where $\langle dW/dt\rangle $ is the averaged (with respect
to the period of the radiation field) energy loss rate. We assume that the frequency of the radiation field
$\omega _{0}$ is larger than the effective frequency of the pairwise Coulomb collisions $\nu _{\mathrm{eff}}$.
Also assuming that in the low--velocity limit the energy loss of the ion on the collective plasma excitations
is negligible and is mainly determined by the Coulomb collisions we obtain $\langle dW/dt\rangle \sim %
\nu _{\mathrm{eff}} W$. On the other hand $\nu _{\mathrm{eff}}\sim 1/v_{\mathrm{eff}}^{3}$, where $v_{\mathrm{eff}}$
is the averaged relative velocity of the colliding particles. At $v<v_{\mathrm{th}}$ and for vanishing radiation
field $v_{\mathrm{eff}} \simeq v_{\mathrm{th}}$. However, in the presence of the radiation field the averaged
relative velocity of the collisions is $v_{\mathrm{eff}} \simeq (v_{\mathrm{th}}^{2} +v_{E}^{2})^{1/2}$
and increases with the intensity of the laser field. Thus the effective collision frequency $\nu _{\mathrm{eff}}$
and hence the stopping power of the ion are reduced with increasing intensity of the radiation field.

At the end of this section we consider a practical example. Let us consider the stopping of the $\alpha $--particles
in the corona of the laser plasma. Although the thermonuclear reactions mainly occur far below the critical surface
the stopping length of the $\alpha $--particles is larger than the characteristic length scale of plasma inhomogeneity
and some part of the $\alpha$--particles transfer the energy to the plasma corona before they reach the critical
surface \citep{max82}. In the vicinity of the plasma critical density the intensity of the radiation field is very
large and the stopping capacity of the plasma may be strongly reduced. In this example the typical temperature is
$T=10$~keV and therefore $v_{\alpha }/v_{\mathrm{th}}=0.22$ ($E_{\alpha }=M_{\alpha }V_{\alpha }^{2}/2=3.5$~MeV, where
$E_{\alpha }$, $M_{\alpha }$, $v_{\alpha }$ are the energy, the mass and the velocity of the $\alpha $--particles).
For $\lambda _{0}=0.5$~$\mu$m, $W_{L}=2\times 10^{17}$~W/cm$^{2}$, and $\omega_{0}=\omega _{p}\sqrt{2}$ (the plasma
density is $n_{0}=n_{c}/2$, where $n_{c}$ is the plasma critical density) we find $a\simeq \lambda _{\mathrm{D}}$.
In this parameter regime the radiation field reduces the SP of the $\alpha $--particles by 20~\%.

\section{Energy loss of fast ions}
\label{sec:3}

In this section we consider the energy loss of a fast heavy ion moving in a fully degenerate plasma (which means
that the partially degenerate case could be postponed to a further presentation) in the presence
of a radiation field. The longitudinal dielectric function of the degenerated electron gas is determined by
Lindhard's expression \citep{lin54,lin64}. However, here we consider the simplest model of the dielectric
function of a jellium. Previously a plasmon--pole approximation to $\varepsilon (k,\omega )$ for an electron gas
was used for calculation of the SP \citep{bas82,deu95,ner00}. In order to get easily obtainable analytical results,
\citet{bas82} employed a simplified form that exhibits collective and single--particle effects
\begin{equation}
\Ima\frac{-1}{\varepsilon (k,\omega )}=\pi \omega _{p}^{2}\frac{|\omega |}{%
\omega }\left[ \delta \left( \omega ^{2}-\omega _{p}^{2}\right)
H(k_{c}-k)+\delta \left( \omega ^{2}-\omega _{k}^{2}\right) H(k-k_{c})\right] ,
\label{eq:23}
\end{equation}%
where $H(x)$ is the Heaviside unit--step function, $\omega _{k}=\hbar k^{2}/2m $, $k_{c}=(2m\omega _{p}/\hbar )^{1/2}$,
and $\omega _{p}$ is the plasma frequency. The cutoff parameter $k_{c}$ is determined by equating the arguments of
the two delta--functions in Eq.~\eqref{eq:23} at $k=k_{c}$. The first term in Eq.~\eqref{eq:23} describes the response
due to nondispersive plasmon excitation in the region $k<k_{c}$, while the second term describes free--electron recoil
in the range $k>k_{c}$ (single--particle excitations). Note that this approximate dielectric function satisfies at
arbitrary $k$ the usual frequency sum rule \citep{bas82,deu95,ner00}.

In contrast to the previous section we consider here the fast projectile ion with $v\gtrsim v_{c}$ (where $v_{c}=\omega_{p}/k_{c}%
=(\hbar\omega_{p}/2m)^{1/2}$) which justifies the approximation \eqref{eq:23} valid only in this specific case \citep{bas82}.

It is constructive to consider first the case of a weak radiation field ($k_{c}a<1$) at arbitrary angle $\vartheta $
between $\mathbf{v}$ and $\mathbf{a}$. In this case the SP is determined by Eq.~\eqref{eq:25}, where the field--free
SP $S_{B}$ in the high--velocity limit is given by \citep{lin54,lin64,deu86,deu95}
\begin{equation}
S_{B}= \frac{Z^{2}e^{2}\omega _{p}^{2}}{v^{2}}\ln \left( \frac{2mv^{2}}{%
\hbar \omega _{p}}\right) .
\label{eq:26}
\end{equation}

Inserting Eq.~\eqref{eq:23} into \eqref{eq:25} for the stopping power we obtain
\begin{equation}
S=\frac{2Z^{2}\Sigma_{0}}{\lambda^{2}}\left\{\ln \lambda
+\frac{(k_{c}a)^{2}}{4}\left[ \Phi_{1}(\lambda ,\gamma )+ \frac{1}{2}
\Phi_{2}(\lambda ,\gamma ) \sin^{2}\vartheta \right] \right\} ,
\label{eq:27}
\end{equation}%
where $\Sigma_{0} =e^{2}k_{c}^{2} =2\hbar\omega_{p}/a_{0}$, $a_{0}$ is the Bohr radius, $\Phi_{1}=\Phi_{1c}+%
\Phi _{1s}$, $\Phi_{2} =\Phi_{2c} +\Phi _{2s}$, $\lambda =v/v_{c}$, $\gamma =\omega _{p}/\omega _{0}<1$. Also
\begin{equation}
\Phi _{1c}(\lambda ,\gamma )=\frac{1}{2\lambda ^{2}}\left[\frac{6}{\gamma^{2}} \ln\lambda
+ \left(\frac{1}{\gamma }+1\right) ^{3}\ln \frac{\gamma }{1+\gamma }%
-\left( \frac{1}{\gamma }-1\right) ^{3}\ln \frac{\gamma }{1-\gamma }\right] ,
\label{eq:28}
\end{equation}%
\begin{equation}
\Phi _{2c}(\lambda ,\gamma )=-3\left[ \Phi _{1c}(\lambda ,\gamma )+\frac{1}{2\gamma ^{2}\lambda ^{2}}\right] ,
\label{eq:29}
\end{equation}%
\begin{eqnarray}
&&\Phi _{1s}(\lambda ,\gamma )=\frac{1}{4\lambda ^{2}}\left[ \frac{1}{2}%
\left( \beta _{1}^{2}+\eta _{1}^{2}-\alpha _{1}^{2}-\delta _{1}^{2}\right) +%
\frac{3}{\gamma }\left( \beta _{1}+\delta _{1}-\alpha _{1}-\eta _{1}\right)
\right.   \label{eq:30} \\
&&\left. -\frac{1}{\gamma ^{3}}\left( \frac{1}{\beta _{1}}-\frac{1}{\alpha
_{1}}-\frac{1}{\eta _{1}}+\frac{1}{\delta _{1}}\right) +\frac{3}{\gamma ^{2}}%
\ln \frac{\beta _{1}\eta _{1}}{\alpha _{1}\delta _{1}}+1-\lambda ^{4}\right] \;,  \nonumber
\end{eqnarray}%
\begin{eqnarray}
&&\Phi _{2s}(\lambda ,\gamma )=\frac{\beta _{1}-\alpha _{1}}{4}\left( 1-%
\frac{9}{\gamma \lambda ^{2}}\right) +\frac{\eta _{1}-\delta _{1}}{4}\left(
1+\frac{9}{\gamma \lambda ^{2}}\right)   \nonumber  \\
&&-\frac{3}{8\lambda ^{2}}\left( \beta
_{1}^{2}+\eta _{1}^{2}-\alpha _{1}^{2}-\delta _{1}^{2}\right)
+\frac{3}{4\gamma ^{3}\lambda ^{2}}\left( \frac{1}{\beta _{1}}-\frac{1}{%
\alpha _{1}}-\frac{1}{\eta _{1}}+\frac{1}{\delta _{1}}\right)   \label{eq:31}  \\
&&+\frac{1}{4\gamma }\left( \ln \frac{\beta _{1}\delta _{1}}{\alpha _{1}\eta _{1}}-\frac{%
9}{\gamma \lambda ^{2}}\ln \frac{\beta _{1}\eta _{1}}{\alpha _{1}\delta _{1}}%
\right) +\frac{1}{4}\left( 1-\frac{1}{\lambda ^{2}}\right) \left( \lambda
^{2}+3\right) ,  \nonumber
\end{eqnarray}%
\begin{eqnarray}
&&\left(
\begin{array}{c}
{\alpha _{n}} \\
{\eta _{n}}%
\end{array}%
\right) =\max \left[ \left( \frac{\lambda }{2}-\sqrt{\frac{\lambda ^{2}}{4}%
\mp \frac{n}{\gamma }}\right) ^{2};\;1\right] ,  \label{eq:32}  \\
&&\left(
\begin{array}{c}
{\beta _{n}} \\
{\delta _{n}}%
\end{array}%
\right) =\left( \frac{\lambda }{2}+\sqrt{\frac{\lambda ^{2}}{4}\mp \frac{n}{%
\gamma }}\right) ^{2} .  \nonumber
\end{eqnarray}
In Eq.~\eqref{eq:32} $n$ is a positive integer ($n=1,2,...$). The first term in Eq.~\eqref{eq:27} corresponds
to the field--free SP~\eqref{eq:26} represented in a dimensionless form. The remaining terms proportional to
the intensity of the radiation field ($a^{2}$), describe
the collective (proportional to $\Phi _{1c;\;2c}(\lambda ,\gamma )$) and single--particle (proportional to
$\Phi _{1s;\;2s}(\lambda ,\gamma )$) excitations. It should be noted that the stopping power Eq.~\eqref{eq:27}
is not vanishing only at high--velocities when $\lambda \geqslant 2 /\sqrt{\gamma}$.

Consider next the angular distribution of the SP at low--intensities of the RF. An analysis of the quantity
$P=(S-S_{B})/S_{B}$ (the relative deviation of $S$ from $S_{B}$) for the proton projectile shows that at
moderate velocities ($\lambda \gtrsim 2/\sqrt{\gamma}$) the angular distribution of $P$ has a quadrupole
nature. At $0\leqslant\vartheta \leqslant \vartheta _{0}(\lambda ,\gamma )$, where $\vartheta _{0}(\lambda%
,\gamma )$ is some value of the angle $\vartheta $, the excitation of the waves with the frequencies
$\omega_{0}\pm \omega _{p}$ leads to the additional energy loss. At $\vartheta_{0}(\lambda ,\gamma )\leqslant%
\vartheta \leqslant \pi /2$ the proton energy loss changes sign and the total energy loss decreases. When
the proton moves at angles $\vartheta =\vartheta _{0} (\lambda ,\gamma )$ with respect to the polarization
vector $\mathbf{a}$ the radiation field has no any influence on the SP. However, at very large velocities
($\lambda \gg 2/\sqrt{\gamma}$) the relative deviation $P$ is negative for arbitrary $\vartheta$ and the
radiation field systematically reduces the energy loss of the proton.

Let us now investigate the influence of the intense radiation field on the stopping process when $\mathbf{v}$
is parallel to $\mathbf{a}$. It is expected that the effect of the RF is maximal in this case. From Eqs.~\eqref{eq:9a}
and \eqref{eq:23} we obtain
\begin{eqnarray}
&&S =S_{0}+\frac{Z^{2}\Sigma_{0} }{\lambda ^{2}}\left\{
\sum_{n=1}^{n_{-}}\left( \frac{n}{\gamma }+1\right) J_{n}^{2}\left(Ap_{n}\right) \ln \frac{\lambda
}{n/\gamma +1}\right.   \nonumber  \\
&&-\sum_{n=1}^{n_{+}}\left( \frac{n}{\gamma }-1\right) J_{n}^{2}\left(Aq_{n} \right) \ln \frac{\lambda
}{n/\gamma -1}  \label{eq:33} \\
&&+\frac{1}{2}\sum_{n=1}^{N}\int_{\alpha _{n}(\lambda )}^{\beta _{n}(\lambda
)}\frac{dx}{x^{2}}\left( \frac{n}{\gamma }+x\right) J_{n}^{2}\left(AP_{n}(x)\right)   \nonumber \\
&&\left. -\frac{1}{2}\sum_{n=1}^{\infty }\int_{\delta _{n}(\lambda )}^{\eta
_{n}(\lambda )}\frac{dx}{x^{2}}\left( \frac{n}{\gamma }-x\right)
J_{n}^{2}\left(AQ_{n}(x) \right) \right\} \;,  \nonumber
\end{eqnarray}%
where $A=k_{c}a$, $P_{n}(x)= (1/\lambda )(n/\gamma +x)$, $Q_{n}(x)= (1/\lambda )(n/\gamma -x)$,
$p_{n}=P_{n}(1)$, $q_{n}=Q_{n}(1)$, and
\begin{equation}
S_{0}=\frac{Z^{2}\Sigma_{0} }{\lambda ^{2}}\left[ J_{0}^{2}\left( \frac{%
A}{\lambda }\right) \ln \lambda +\frac{1}{2}\int_{1/\lambda }^{\lambda }%
\frac{dx}{x}J_{0}^{2}\left( Ax\right) \right]
\label{eq:34}
\end{equation}
is the SP without emission or absorption of the photons. Also we have introduced the notations
\begin{eqnarray}
&&n_{\pm }=\Int \left( \frac{k_{c}v\pm \omega _{p}}{\omega _{0}}\right) =\Int \left[
\gamma \left(\lambda \pm 1\right) \right] ,  \label{eq:35} \\
&&N=\Int \left(\frac{mv^{2}}{2\hbar \omega _{0}}\right) =\Int\left( \frac{\gamma \lambda ^{2}}{%
4}\right) , \nonumber
\end{eqnarray}%
where $\Int (x)$ is the integer part of $x$. The quantities $\alpha_{n}(\lambda )$, $\beta_{n}(\lambda )$,
$\delta_{n}(\lambda )$, $\eta_{n}(\lambda )$ in Eq.~\eqref{eq:33} are determined by Eq.~\eqref{eq:32}.
We note that in Eq.~\eqref{eq:33} the terms involving $n_{\pm}$ and $N$ photons are not vanishing at
$\lambda \geqslant 1/\gamma \mp 1$ and $\lambda \geqslant 2/\sqrt{\gamma}$, respectively. Similarly
the SP \eqref{eq:34} is not vanishing at $\lambda \geqslant 1$.

\begin{figure*}[tbp]
\centering{
\includegraphics[width=75mm]{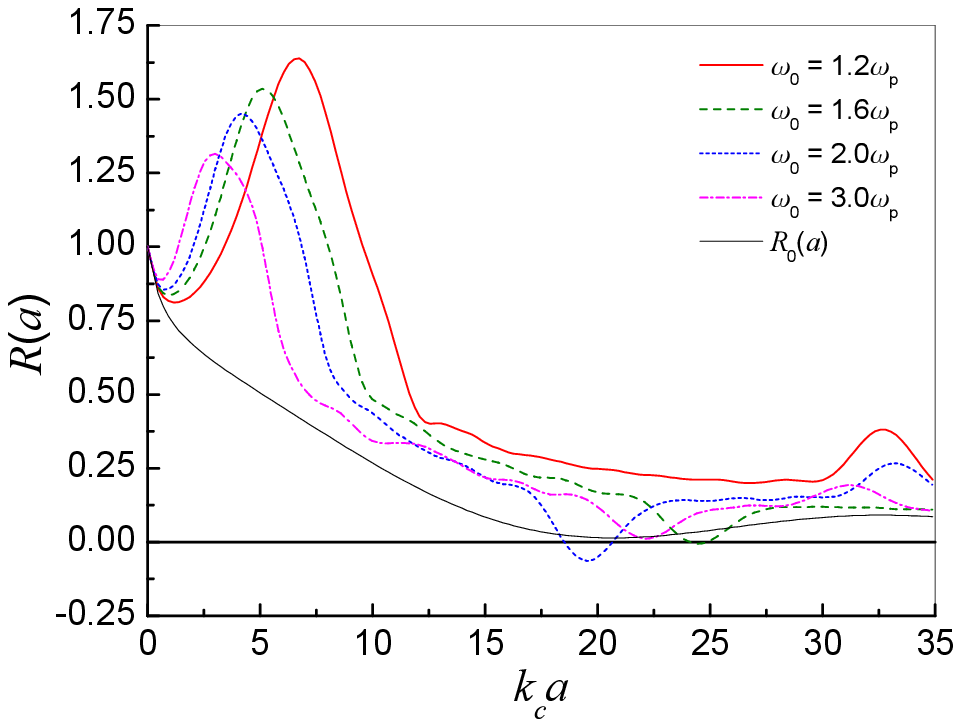}
\includegraphics[width=75mm]{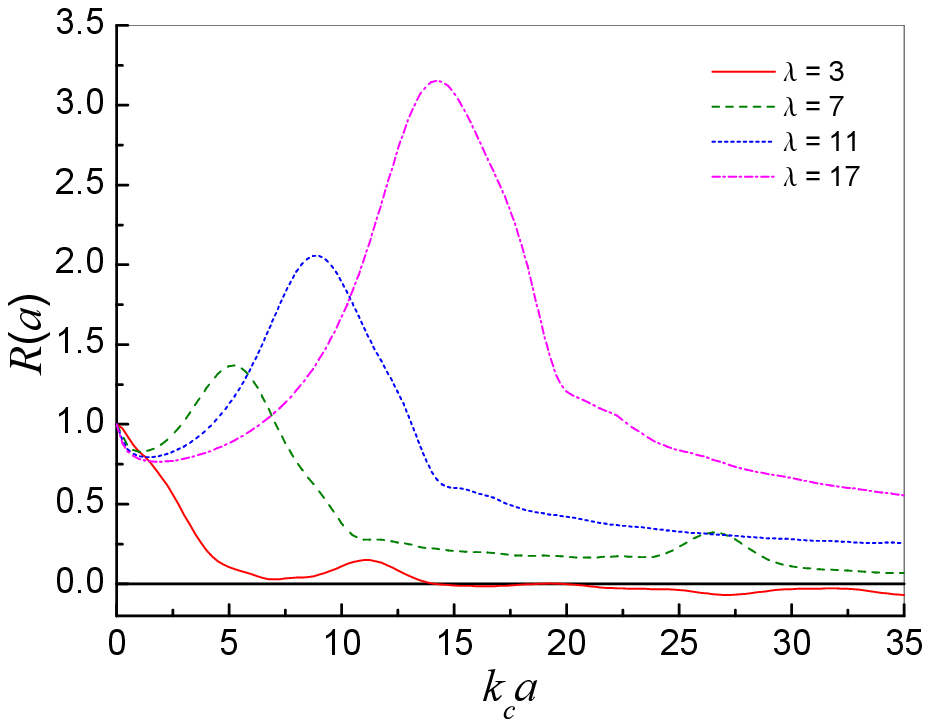}}
\caption{(\textbf{Left panel}) the ratio $R(a)=S(a)/S_{B}$ as a function of dimensionless quantity $k_{c}a$ at $v=8.6v_{c}$,
$\omega _{0}=1.2\omega _{p}$ (solid line), $\omega _{0}=1.6\omega _{p}$ (dashed line), $\omega _{0}=2\omega _{p}$
(dotted line), $\omega _{0}=3\omega _{p}$ (dash--dotted line). Thin solid line corresponds to $R_{0}(a)=%
S_{0}(a)/S_{B}$ (see Eq.~\eqref{eq:34}). (\textbf{Right panel}) same as in left panel but at $\omega _{0}=1.2\omega _{p}$,
$v=3v_{c}$ (solid line), $v=7v_{c}$ (dashed line), $v=11v_{c}$ (dotted line), $v=17v_{c}$ (dash--dotted line).}
\label{fig:2}
\end{figure*}

The first term in Eq.~\eqref{eq:34} describes the collective excitations while the second term corresponds to
the single--particle excitations. From Eq.~\eqref{eq:34} it is seen that $S_{0}$ oscillates with the intensity
of the laser field. However, the radiation field suppresses the excitation of the collective and the single--particle
modes and the SP $S_{0}$ is less than the field--free SP $S_{B}$. As follows from Eq.~\eqref{eq:34} at high--intensities
of the RF the SP $S_{0}$ is close to zero when $A/\lambda \simeq \mu_{m}$ (or alternatively at $\gamma (v_{E}/v)%
\simeq \mu_{m}$) with $m =1,2,\ldots$, where $\mu_{m}$ are the zeros of the Bessel function $J_{0}(\mu_{m})=0$
($\mu_{1}=2.4$, $\mu_{2}=5.52$, $\mu_{3}=8.63$\ldots). Then the energy loss of the ion is mainly determined by
the other terms in Eq.~\eqref{eq:33} and is stipulated by excitation of plasma waves with frequencies $n\omega_{0}%
\pm\omega_{p}$. The first and the last pairs of terms in Eq.~\eqref{eq:33} describe the excitation of the collective
and single--particle modes, respectively, with emission or absorption several photons. The number of photons
($n_{\pm }$, $N$) involved in the process of the inelastic interaction are determined by the energy--momentum
conservations (see the arguments of the delta--functions in the dielectric function \eqref{eq:23}).

The results of the numerical evaluation of the SP (Eqs.~\eqref{eq:33} and \eqref{eq:34}) are shown in Figure~\ref{fig:2},
where the ratio $R(a)=S(a)/S_{B}$ is plotted as a function of the laser field intensity ($k_{c}a=5.38W_{L}^{1/2}\omega%
_{0}^{-2}r_{s}^{-3/4}$, where $r_{s}$ is the Wigner--Seitz density parameter and $W_{L}$ and $\omega _{0}$ are measured
in units $10^{15}$~W/cm$^{2}$ and $10^{16}$~sec$^{-1}$, respectively). For instance, for Al target with $r_{s}=2.07$,
$\hbar\omega_{p}=15.5$~eV, and $v_{c} =1.2\times 10^{8}$~cm/sec. From Figure~\ref{fig:2} it is seen that the SP exceeds the
field--free SP and may change sign due to plasma irradiation by intense ($k_{c}a\gg 1$) laser field. Similar properties
of the SP has been obtained previously for a classical plasma \citep{ner99}. However, due to the higher density of the
degenerate electrons (in metals typically $n_{0}\sim 10^{23}$~cm$^{-3}$) the acceleration rate of the projectile particle
is larger than similar rate in the case of a classical plasma. The acceleration effect occurs at $v_{E}/v\simeq \mu_{m}/\gamma$
(with $m=1,2,\ldots$) when the SP $S_{0}$ nearly vanishes. It should be noted that in the laser irradiated plasma a parametric
instability is expected \citep{sil73} with an increment increasing with the intensity of the radiation field. This restricts
the possible acceleration time with stronger condition than in the case of a classical plasma. Finally, let us note that
the effect of the enhancement of the SP of an ion moving in a laser irradiated plasma is intensified at smaller frequency
(Fig.~\ref{fig:2}, left panel) of the radiation field ($\omega _{0}\simeq \omega _{p}$ but $\omega _{0}>\omega _{p}$) or
at larger incident kinetic energy of the projectile ion (Fig.~\ref{fig:2}, right panel) when the numbers $n_{\pm}$ and $N$
of the photons involved in the inelastic interaction process are strongly increased (Eq.~\eqref{eq:35}).

\section{Summary}
\label{sec:sum}

In this paper, within RPA we have investigated the energy loss of a heavy point--like ion moving in a laser irradiated plasma.
In the course of this study, we derived a general expression for the SP which has been also simplified in the limit of weak
RF. As in the field--free case, the SP in a laser irradiated plasma is completely determined by the dielectric function of the
plasma. We have considered two somewhat distinct cases of the slow-- and high--velocity ion moving in a classical and fully
degenerate electron plasma, respectively. At low--velocities the RF leads to the strong decrease of the energy loss. Physically,
this is due to the strong reduction of the effective frequency of the pairwise Coulomb collisions between projectile ion
and the plasma electrons. At high velocities the RF may strongly increase the SP. This effect is more pronounced when the laser
frequency approaches the plasma frequency in agreement with PIC simulations \citep{hu11}. Moreover, at high--velocities and in the
presence of the intense RF an ion projectile energy gain is expected when the quiver velocity of the plasma electrons exceeds the
ion velocity. The analysis presented above can in principle be extended to the case of a partially degenerate plasma as well as to
the case of light ion projectiles and also electrons and positrons when the effect of the intense RF on the ion cannot be neglected
anymore. We intend to address these issues in our forthcoming investigations.

\section*{Acknowledgments}
The work of H.B.N. has been partially supported by the State Committee of Science of Armenian Ministry of
Higher Education and Science (Project No.~11-1c317).

\end{document}